\documentclass[preprint]{aastex}
\begin{document}
\def\etal{{\it et al.\/}}
\def\cf{{\it cf.\/}}
\def\ie{{\it i.e.\/}}
\def\eg{{\it e.g.\/}}

\title{A simple, stringent test on the nature of GRBs' progenitors}
\author{{\bf Mario Vietri}}
\affil{
Universit\`a di Roma 3, Via della Vasca Navale 84, 00147 Roma, \\
Italy, E-mail: vietri@corelli.fis.uniroma3.it \\
}

\begin{abstract}
I show here that the time--delay between the gamma ray burst proper
and the onset of the afterglow is a sensitive function of the
surrounding density, going from $\approx 10\; s$ for typical disk
densities to several hours for IGM densities. Since
bulk Lorenz factor, explosion energy and environment density can be 
simultaneously determined from observations of the time--delay and of the
afterglow, afterglow observations alone may establish which fraction, if 
any, of all 
bursts resides inside galaxies, in their outer haloes, or outright in the IGM 
medium, subjecting the NS/NS merger hypothesis to a definitive test. 
If bursts are due to collapse of massive stars, no
matter how short the burst, the onset of the afterglow should be immediate.
\end{abstract}


\section{Introduction}

There are currently three favored models for the formation of GRBs:
mergers of neutron star binaries (Narayan, Paczynski and Piran 1992),
collapsars (or hypernovae, Woosley 1993, Paczynski 1998), and
SupraNovae (Vietri and Stella 1998, 1999), but a clearcut test to which
each of these could be subjected is still missing. This is so because
the details of the fireball model are to a large extent independent of the
details of the energy deposition mechanism. The situation has also a
paradoxical side: even if we managed to observe the event that deposits
the energy, the enormous expected optical depths (exceeding $10^{10}$)
would prevent us from deriving any information on the nature of the
source. Gravitational waves can probably provide a signature of at least
some of the scenarios, but detection of these from these distant sources
does not look like an immediate prospect. 

For these reasons, many authors (B\"ottcher \etal 1999,
M\'esz\'aros and Rees 1998) have proposed that interaction of the burst
with surrounding material might provide detectable signatures of
star--forming environments. However, the possible discovery of
iron emission lines (Piro \etal, 1999, Yoshida \etal, 1999) from two
bursts has highlighted the weak side of this possibility, \ie, that
any conclusion is subject to the usual uncertainties associated with
modeling complex and incomplete sets of observations (Lazzati \etal,
1999, Vietri \etal, 1999). I show in this
paper that a surprisingly simple and robust test exists, for both the
neutron binary merger and the collapsar scenarios, resulting from a
trivial, but so far overlooked property of fireballs.

\section{Time delays}

I consider below spherically symmetric fireballs. Since I am interested in
phases of the evolution when the ejecta are still highly relativistic, 
the corrections due to the sideways expansion of a {\it beamed} model 
are expected to be small. The test is provided by the delay
of the onset of the afterglow from the end of the burst proper.
The burst duration is due exclusively to the total lifetime of the
source, a quantity which we cannot currently predict, but which, given the 
large luminosities involved, is observed to lie within a hundred seconds. The 
radius at which the energy release occurs, at the contrary, is completely fixed 
by the requirement that the source geometry does not filter out the
millisecond--scale variability which is so often observed: we must have
$R_{sh} \approx 10^{13}\; cm$ (Ruderman 1975, Sari and Piran 1997).
At these distances, the relative kinetic motion of the dishomogeneous
wind is dissipated. The afterglow proper begins instead at a distance
which is fixed solely by the density of the surrounding medium, and the
overall properties of the wind: 
\begin{equation}
R_{ag} = 4\times 10^{16}\; cm
\left(\frac{E}{10^{53}\;erg}\right)^{1/3}
\left(\frac{1\; cm^{-3}}{n}\right)^{1/3}
\left(\frac{300}{\gamma}\right)^{2/3}
\end{equation}
where $n$ is the number density of surrounding matter, and $\gamma$ the bulk
Lorenz factor. At this radius, the inertia of swept--up matter becomes
comparable to that of the ejecta, and slowdown begins. From this point on,
the X--ray luminosity begins decreasing as $t^{-\alpha}$, with $\alpha =
0.7-2.2$, the complete range observed so far. The beginning of this
easily recognizable phase occurs a time $t_d$ after the end of the burst,
with
\begin{equation}
\label{td}
t_d = \frac{R_{ag}-R_{sh}}{\gamma^2 c} \approx
\frac{R_{ag}}{\gamma^2 c} = 15\; s
\left(\frac{E}{10^{53}\;erg}\right)^{1/3}
\left(\frac{1\; cm^{-3}}{n}\right)^{1/3}
\left(\frac{300}{\gamma}\right)^{8/3}\;.
\end{equation}

The time--delay is a function of the product $n \gamma^8$, so that one
may fear that large variations of $n$ may be masked or mimicked by much 
smaller variations in $\gamma$; this degeneracy will be lifted shortly,
but it should be remarked that that the ejecta Lorenz factor is limited 
from below to $\gamma \ga 30$ (M\`esz\`aros, Laguna and Rees 1993), and 
it is unlikely to exceed $\gamma \approx 300$ by much. At the same time, 
considering for the moment $\gamma$ fixed, we see that
the time delay $t_d$ is a moderately sensitive function of the
surrounding density: for a density typical of the interstellar medium
in a galactic disk, $n \approx 1\; cm^{-3}$, clearly $t_d \approx 10\;
s$. For long bursts, and in particular for the whole subclass of bursts
accessible to BeppoSAX, this time--delay is comparable with the burst
duration $t_b$, and we would thus expect a superposition of the last
parts of the burst and the beginning of the afterglow. 
However, the time--delay acquires interesting and measurable values when
the burst takes place outside a disk: in a galactic halo, for which a
baryonic representative density is around $n \approx 10^{-4} \; cm^{-3}$,
I find $t_d \approx 5$ minutes. In the case of the intergalactic medium,
for which $n \approx 10^{-8}\; cm^{-3}$, I find $t_d \approx 4\; h$.

So we see that the expected range of variation for $\gamma^8$, about $8$
orders of magnitude, is similar to the expected range for $n$: if we 
find an independent way of fixing $\gamma$, then we may use the time--delay 
as a measure of whether the environment density really spans eight orders
of magnitude, and thus, in the end, of where the bursts dwell. That 
afterglow observations could fix the environment parameters had been 
noticed already in the literature; what is new here is the realization
that the difference in environmental densities controls the time--delay
as much as the ejecta Lorenz factor, and thus that the time--delay can be 
turned into a sensitive measure of the bursts' whereabouts. 

What happens between the end of the burst proper and the onset of the
afterglow? Electrons accelerated at internal shocks, both by
quasi--thermal and Fermi processes, cool much faster than the
hydrodynamical expansion time--scale, (Sari, Narayan and Piran 1996). However, 
some electrons will be accelerated at the forward shock which separates
the surrounding medium from shocked matter already raked up by the ejecta.
Since the rate at which matter is raked up is roughly constant, we can take
as a first approximation that the total emitted power is a constant, for
which an upper limit can be found as follows. The shock slows down when
the total swept up mass (times $\gamma$) equals the ejecta mass. Within
this time interval, then, about half of the total energy is transfered
to the swept up mass; if this were highly radiative, then the
total emitted power would be of order $\dot{E} \approx \frac{E}{t_d}$
which is much smaller (by the factor $t_{b}/\xi t_d$, where $\xi
\approx 0.1$ is the ratio of relative to bulk kinetic energy of the wind)
than the burst luminosity. However, this is an upper limit, because, though 
electrons are highly radiative, this estimate implies
that they are able to radiate away also all of the thermal energy of the
protons. Since it is generally assumed (and it can be checked in the
observations of the afterglow of GRB 970508, Frail, Waxman and Kulkarni
2000) that only a fraction $\epsilon_{eq} \approx 0.1$ of the protons'
energy can be transferred from the protons to the electrons, this
assumption is excessive by the same factor, $\epsilon_{eq}$. 
The exact luminosity radiated during this coasting
phase is difficult to predict exactly, since it depends critically upon
rather uncertain plasma factors, like the efficiency $\epsilon_{eq}$
with which energy is tranferred from protons to electrons, the efficiency
$\epsilon_B$ with which a near--equipartition magnetic field is built up,
and the relative efficiency of quasi--thermal to Fermi processes.
However, for the present aim, this exact value is unnecessary: suffice it
to say that a reasonable expectation is that
\begin{equation}
L \la \frac{\epsilon_{eq}E}{t_d}\;.
\end{equation}

What happens after shock deceleration has been described, for different
values of the parameters, by M\'esz\'aros, Laguna and Rees (1993).
Basically, a reverse shock will propagate at moderate Lorenz factors
backward into the ejecta, dissipating an appreciable fraction of the
kinetic energy of the shell within a timescale $R_{ag}/\gamma^2 c$,
which is the apparent time in which the shock crosses the ejecta
shell, $\approx t_d$. Thus the maximum expected power
is exactly a fraction  $t_{b}/\xi t_d$ below that of the burst.
Notice that, in the previous section, this was only an upper limit.
This internal shock provides a new, second phase of emission, of
total duration comparable to the light crossing time of the ejecta shell,
$t_d$. 

After this peak luminosity we then expect the usual afterglow
($\propto t^{-\alpha}$) to begin. 
A qualitative description of this can be seen in Fig. 1.
During the afterglow, emission is due to synchrotron processes (M\`esz\`aros
and Rees 1997), as evidenced by the simulataneous spectrum of GRB 970508
that fits theoretical expectations so nicely (Galama \etal, 1998) and by
detection of polarization in the optcial spectrum of GRB 990510 (Covino
\etal, 1999, Wijers \etal, 1999). In this case, it is well--known that the
synchrotron peak frequency $\nu_m$  is given by (Waxman 1997a)
\begin{equation}
\nu_m = 6\times 10^{14}\; Hz \left(\frac{1+z}{2}\right)^{1/2} 
(\epsilon_{eq}/0.2)^2 (\epsilon_B/0.1)^{1/2} E_{53}^{1/2} t_{day}^{-3/2}
\left(\frac{n}{1\; cm^{-3}}\right)^{1/2}
\left(\frac{\gamma}{190}\right)^4
\end{equation}
where $E_{53}$ is the explosion energy in units of $10^{53}\; erg$, 
$t_{day}$ is the time of observations in units of days after the beginning of 
the afterglow, and $\gamma$ is the initial bulk Lorentz factor before 
the inertia of the swept--up matter begins slowdown. Using Eq. \ref{td}, it is
convenient to rewrite this as a function of the time--delay $t_d$:
\begin{equation}
\label{num}
\nu_m = 3\times 10^{15}\; Hz \left(\frac{1+z}{2}\right)^{1/2} 
(\epsilon_{eq}/0.2)^2 (\epsilon_B/0.1)^{1/2} E_{53} t_{day}^{-3/2}
\left(\frac{15\; s}{t_d}\right)^{3/2}\;.
\end{equation}
As an example, for the time--delays discussed above, 
ten minutes after the beginning of the afterglow we expect 
the peak frequency to be $h \nu_m = 22 \; keV $ for the disk ISM, 
$250 \; eV$ for the halo, and $1\; eV$ for the IGM.

On the other hand, the intensity at $\nu_m$ is independent of the bulk Lorenz 
factor (Waxman 1997b):
\begin{equation}
\label{fnum}
F(\nu_m) = 10\; mJy \left(\frac{\epsilon_B}{0.1}\right)^{1/2}
\frac{2}{1+z} \left(\frac{1-1/\sqrt{2}}{1-1/\sqrt{1+z}}
\right)^2 \left(\frac{n}{1\; cm^{-3}}\right)^{1/2} E_{53}^{1/2}\;.
\end{equation}
but is a reasonably steep function of the environmental density. 
Eq. \ref{td},\ref{num}, \ref{fnum} allow the simultaneous solution for the 
three unknown parameters $E$, $n$ and $\gamma$, which are the only ones 
potentially varying by several orders of magnitude. In this way, we can lift
the degeneracy present in the factor $n\gamma^8$. Basically, given a
time--delay, a time $t_{day}$ after the new peak we may locate the
afterglow peak at a frequency given by Eq. \ref{num}, and then the
afterglow intensity (Eq. \ref{fnum}) will establish whether the time--delay
is given a much smaller value of $n$ or to a marginally smaller value
of $\gamma$. In bands like the X--ray, if the time--delay is due to
$n \ll 1\; cm^{-3}$, we basically do not expect any detectable flux,
which might however be detectable in the optical/UV, while the afterglow
is certainly detectable in the optical if a smaller value of $\gamma$
accounts for the time--delay. For this reason, experiments like SWIFT,
which will provide simultaneous coverage in both optical/UV and in the
X--ray, are of fundamental importance to interpret time--delays. 

Another interesting application of Eq. 1 occurs for the opposite case,
in which the surrounding material is that of a pre--existing wind from
the progenitor system. In this case, using $n = \dot{M}/4\pi r^2 m_p
v_\infty$, I find
\begin{equation}
R_{ag} = \frac{E v_\infty}{\dot{M}\gamma^2 c^2} = 10^{14}\; cm
\frac{E}{10^{52}\; erg}
\frac{v_\infty}{100\; km\; s^{-1}}
\frac{10^{-6} M_\odot\; yr^{-1}}{\dot{M}}
\left(\frac{300}{\gamma}\right)^{2}
\;.
\end{equation}
From this it can be seen that only exceedingly short bursts, \ie, those
lasting less than $R_{ag}/\gamma^2 c \approx 0.03\; s$, can display a
break in the X--ray emission: all other bursts (\ie, the near totality)
must display an afterglow emission superimposed on the burst itself, and
no break of continuity. 

\section{Discussion}

The relevance of the above arguments to the problem of the identification
of the bursts' progenitors is clear. Bloom, Sigurdsson and Pols
(1999) have shown that there is a well--defined distribution of
expected distances of GRBs from their parent galaxies, if they 
originate from the merger of neutron stars. In particular, they showed
that $\approx 15\%$ of all burst progenitors may be able to escape from
their parent galaxy altogether, and that about $50\%$ of all bursts
will be located more than $8\; kpc$ from their site of origin, nearly in
the galactic halo. This feature is potentially verifiable
also with optical investigations, but we know that even the most
powerful ground--based telescopes are often inadequate to find the
host galaxy, while HST searches, which up to now have scored a remarkable
$5/6$ success rate (Fruchter 2000), are slow, cumbersome, and, for some
of the most distant bursts, cannot go deep enough in the luminosity
function. The test I propose is instead simple, and a statistically meaningful 
database will be secured in the near future, since several space missions 
(AXAF, Hete II, Integral, XMM, SWIFT) will observe hundreds of bursts
for several hours after the burst. Especially important among them is 
SWIFT, which will be responsible for most of these observations, and will
provide continuous coverage of each detected bursts in
both X--ray and UV/optical wavebands. From this database of perhaps $300$
bursts observed continuously over several hours after the burst, we may
expect $\approx 50$ bursts located in the IGM medium, and another $\approx
100$ located in the haloes of their parent galaxies. Thus, within this
scenario we expect about a third of all bursts to have delays of several
minutes, and a few tens to have delays of a few hours. The flux should then
peak again to a factor $\approx t_{b}/\xi t_d$ below that of the burst
proper.

The very same test can also be used {\it a converso} to check whether
bursts are surrounded by the massive winds expected in all scenarios
involving massive stars.
In particular, it is expected that no time--delay will ever be observed,
except perhaps for the shortest bursts, $t_{b} \ll 0.03\;s$.

One may wonder what happens when only incomplete data are available, \ie,
when the burst disappears shortly after $t_b$ and is re--observed when
already in the afterglow regime. This is the case of BeppoSAX which,
because of instrumentational limitations, is incapable of providing
significant information on the key silent period described here. It should
be notice that the above estimates for the luminosity ($L_b$ at $t_b$, and
$L_b t_d/\xi t_d$ at $2 t_d$) fall precisely on a $t^{-1}$ slope. So, if
further observations of the afterglow find an afterglow slope close to
$-1$, extrapolation of the afterglow luminosity to early time will match
closely with the burst luminosity, regardless of the existence of the 
silent period. Thus, when only incomplete data are available, only
bursts with slopes significantly different from $-1$ can be helpful in
deciding whether a silent period exists. 

A corollary of the above tests comes from the observations of flares
in the radio lightcurves. In fact, these flares are expected only from
afterglow sources with radii smaller than $R_c$ (Goodman 1997), where
\begin{equation}
R_c = 10^{17}\; cm \frac{\nu_{10}^{6/5}}{d_{sc,kpc}h_{75}}
\left(\frac{SM}{10^{-2.5} m^{-20/3} kpc}\right)^{-3/5}\;.
\end{equation}
Here $\nu_{10}$ is the observing frequency in units of $10\; GHz$,
$d_{sc,kpc}$ the distance of the Galactic scattering medium, assumed
to be a simple screen, in units of $1\; kpc$, and $SM$ is the Galactic
scattering measure, scaled in terms of a typical Galactic value. It
can be seen from the above that $R_c \ll R_{ag}$ every time that the
burst is not located inside a galactic disk. Thus, for the $\approx
150$ sources mentioned above, which are expected to show measurable
time--delays, we expect {\it a fortiori} no radio flares. Radio flares
are instead allowed if bursts originate from collapsars.                                          

In summary, I have shown here that the X--ray luminosity has a silent
period between the burst and the power--law afterglow, of duration
\begin{equation}
t_d = \left\{ 
\begin{array}{ll}
15 \; s & \mbox{ISM} \\
5 \; min & \mbox{galactic halo} \\
4\; h & \mbox{IGM}
\end{array}
\right.
\end{equation}
after which we expect a resurgence in the bolometric flux to a level a factor
$\approx t_{b}/\xi t_d$ below that of the burst proper, from which
the easily recognizable, power--law afterglow begins. Afterglow intensity 
at peak frequency depends only on the environmental density and total
explosion energy, while the peak frequency depends on total energy, 
$n$ and $\gamma$, allowing a simultaneous solution which fixes $n$ 
(Eqs. \ref{td}, \ref{num}, \ref{fnum}). 

{}

\newpage
\begin{figure}
\epsscale{0.5}
\plotone{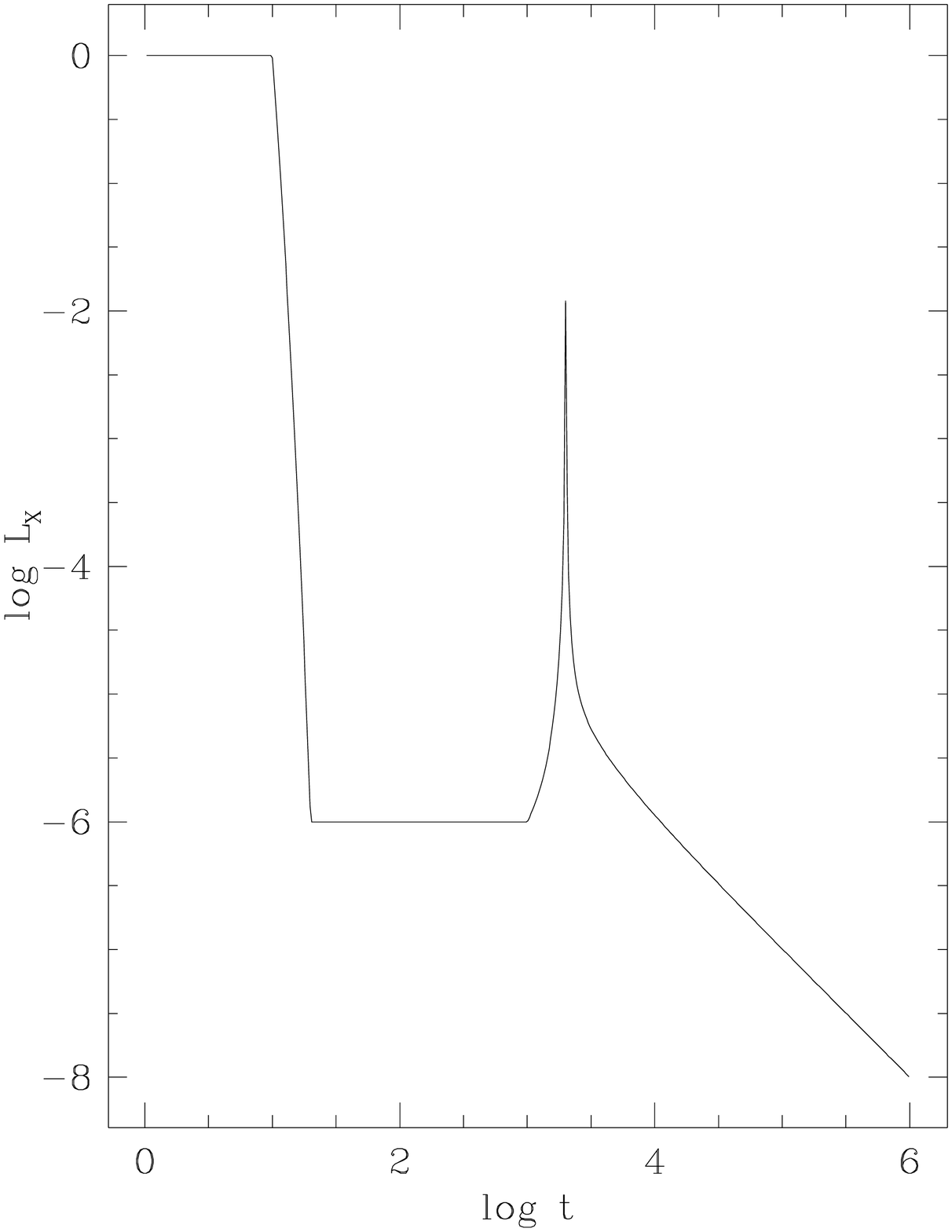}
\caption{Qualitative X--ray lightcurve of burst with measurable 
time--delay between burst proper, and afterglow. During $t_b = 10\; s$, there
is the X--ray-- bright burst; during the next $t_c < t_b$, the electrons
accelerated at the internal shocks cool, and the X--ray luminosity decays to
a small value; for $t < t_d = 10^3\; s$, the X--ray flux is due to the
small fraction of the (small) radiated luminsoity which ends in the X--ray
band: most of the flux will be in the $\gamma$--ray band. When $t \approx
t_d = 10^3\; s$, the forward shock begins decelerating, driving a reverse 
shock into the ejecta; this causes a resurgence of the X--ray luminosity
on a time--scale $\approx t_d$, up to a flux level a factor $t_b/\xi t_d$ 
below that of the burst. At a time $\approx t_d$, the afterglow
proper begins, first driving a reverse shock into the ejecta (which is 
responsible for the new peak), then beginning the the power--law--like
decay. Notice that the normalization of the afterglow (Eq. \ref{num},
\ref{fnum}) is unrelated to that of the peak.}
\end{figure}

\end{document}